# Modeling the Galactic Center Nonthermal Filaments as Magnetized Wakes

Russell B. Dahlburg

*Laboratory for Computational Physics & Fluid Dynamics, Naval Research Laboratory, Washington, DC 20375-5344*

rdahlbur@lcp.nrl.navy.mil

Giorgio Einaudi

*Dipartimento di Fisica e Istituto Nazionale Fisica Materia, Sez. A, Università di Pisa, 56100, Pisa, ITALIA*

einaudi@df.unipi.it

T. N. LaRosa

*Department of Biological and Physical Sciences, Kennesaw State University, 1000 Chastain Rd., Kennesaw, GA 30144*

ted@avatar.kennesaw.edu

Steven N. Shore

*Department of Physics & Astronomy, Indiana University South Bend, 1700 Mishawaka Avenue, South Bend, IN 46634-7111*

sshore@paladin.iusb.edu

## ABSTRACT

We simulate the Galactic Center nonthermal filaments as magnetized wakes formed dynamically from amplification of a weak (tens of $\mu$G) global magnetic field through the interaction of molecular clouds with a Galactic Center wind. One of the key issues in this *cometary model* is the stability of the filament against dynamical disruption. Here we show 2-dimensional MHD simulations for interstellar conditions that are appropriate for the Galactic Center. The structures eventually disrupt through a shear driven nonlinear instability but maintain coherence for lengths up to 100 times their width as observed. The final instability, which destroys the filament through shredding and plasmoid formation, grows quickly in space (and time) and leads to an abrupt end to the structure, in accord with observations. As a by-product, the simulation shows that emission should peak well downstream from the cloud-wind interaction site.

*Subject headings:* Galaxy:Center — nonthermal emission

## 1. Introduction

Continuum radio observations of the Galactic Center reveal a complex environment with many unique structures. Among the most interesting sources are the isolated nonthermal filaments (hereafter NTFs) which are remarkably coherent magnetic structures that extend tens of parsecs (10-60 pc) and maintain widths of only few tenths of parsecs (see Morris 1996, Morris & Serabyn 1996, Lang, Morris & Echevarria 1999, LaRosa et al. 2000). Their magnetic field strengths are estimated to be about 1 mG based on presumed stability of the filaments against ram pressure from ambient cloud collisions (Yusef-Zadeh & Morris 1987a,b). This is three orders of magnitude larger than the general ISM magnetic field. Polarization studies have shown that the magnetic fields of the NTFs are aligned axially within the filaments. At higher resolution, total intensity maps often show the NTFs that appear to be braided (Yusef-Zadeh, Parastaran & Wardle 1997; Lang et al. 1999). Flaring at the ends of the filaments is also seen (Lang, Morris & Echevarria 1999; Lizst & Spiker 1995). Their surface brightness profiles are not uniform along their lengths, often peaking near the midpoint of a filament – where subfilaments appear to overlap – and decreasing smoothly in both directions (LaRosa et al. 2000). Yet this brightness profile is also found in at least one NTF system that shows no braiding (LaRosa, Lazio & Kassim 2001) suggesting this behavior may be a generic feature of the NTF phenomenon.

The isolated NTFs show steep spectral indices. Between 20 and 90 cm, using $S = S_0\nu^p$, values of $-0.6 < p < -0.4$, while above 5 GHZ the spectra are steeper: $p \sim -1.5$ and the NTFs become increasingly more difficult to detect. The spectra are indicative of a curved electron energy distribution. Assuming an initial power-law energy spectrum the observed curvature suggests an age of 2-4$\times 10^4$ years for a filament. This contrasts sharply with observed properties of the *bundled* filaments in the Galactic Center Radio Arc (Yusef-Zadeh, Morris & Chance 1984). The bundled filaments show little substructure, flat spectral indices, and can be detected at frequencies as high as 150 GHz (Sofue, Riech & Matsuo 2000). This likely requires ongoing electron acceleration in these structures.

These filamentary structures appear to be unique to the Galactic Center region and therefore require special conditions that are not present in the general interstellar medium.

It is widely thought that they are tracing a globally ordered magnetic field that permeates the whole inner 100 to 200 parsecs (Serabyn & Morris 1996). However, the origin and stability of such a strong magnetic field over this large scale has been not clearly established (although see Chandran, Cowley, & Morris 2000). Although several suggestions have been made for how electron acceleration could occur in this scenario, quantitative models have yet to be developed.

With these problems in mind, Shore & LaRosa (1999) recently proposed a dynamical model in which the NTFs are explained as wakes resulting from the interaction of a molecular cloud with a Galactic Center wind, analogous to the formation of a comet tail in the solar wind. Assuming that the wind advects magnetic flux, the field cannot penetrate the obstacle (*e.g.* a molecular cloud) and the flow stretches the field around the cloud forming a long, thin magnetotail. This scenario for filament formation has been simulated successfully by Gregori et al. (2001) using a 3-D MHD code. These authors considered a spherical cloud moving transversely to a static magnetic field. They find for an initial ratio of kinetic to magnetic energy, $\beta \equiv P_{gas}/P_{mag} \approx 100$, that the magnetic field is amplified by stretching and compression by a factor $>10^2$ and is structured into flux ropes that merge along the wake's symmetry axis.

The most critical issue yet to be addressed, however, is the stability of the filaments thus produced. In this paper work we numerically simulate the time evolution of a wake flow using an *independent* 2D MHD code. Simulations of comets in the solar wind have been performed by Rauer et al. (1995) using a 3D MHD code. These calculations are relevent to our case in showing how our initial conditions can be produced. Although we suppress the twisting and bending modes that fully 3D models can produce, the spherical blunt body wakes and tails found by Rauer et al. show that the helical modes do not limit the growth of the tail. Large aspect ratios are obtained and the basic current sheet behavior is similar to what we see for a 2D simulation. With our higher spatial resolution code, as we will show, the wake flows are indeed unstable in two dimensions and the instability limits the length of a filament.

## 2. Numerical Method and Initial Conditions

The evolution of a wake flow in a magnetized plasma has been modeled in the context of accelerated streamer flows in the solar corona and solar wind (e.g. Einaudi et al. 1999, 2001; Dahlburg et al. 1998, 2001). Since the physical scenario is strikingly similar to the one invoked by Shore & LaRosa (1999) for the NTFs, we here examine the application of this same simulation technique to the Galactic center. The most significant results of this work is that a wake flow with a sheared magnetic field is unstable and that the slowly moving wake material is accelerated up to some fraction of the driving wind speed. Consider the physical situation depicted in Figure 1. Initially the wake velocity profile is zero at the midpoint and smoothly joins the outside flow with a characteristic width of $\Delta$ which we identify with the thickness of the filament. The magnetic field reverses direction along the symmetry and forms a neutral current sheet.

## 2.1. Equations of the Problem

We start with the nonlinear partial differential equations which govern the behavior of a three-dimensional, compressible, dissipative magnetofluid, written here in a dimensionless form:

$$\frac{\partial \rho}{\partial t} = -\nabla \cdot (\rho \mathbf{v}) \tag{1}$$

$$\rho \left( \frac{\partial \mathbf{v}}{\partial t} + \mathbf{v} \cdot \nabla \mathbf{v} \right) = -\frac{1}{\gamma M^2} \nabla P + \frac{1}{M_A^2} \mathbf{J} \times \mathbf{B} + \frac{1}{R} \nabla \cdot \zeta \tag{2}$$

$$\rho \left( \frac{\partial T}{\partial t} + \mathbf{v} \cdot \nabla T \right) = \frac{\gamma - 1}{\gamma} \left( \frac{\partial P}{\partial t} + \mathbf{v} \cdot \nabla P \right) + \frac{M^2}{R} (\gamma - 1) \, \zeta_{ij} e_{ij}$$
$$+ \frac{M^2}{R_m M_A^2} (\gamma - 1) (\nabla \times \mathbf{B})^2 + \frac{1}{R \, Pr} \nabla^2 T \tag{3}$$

$$\frac{\partial \mathbf{B}}{\partial t} = \nabla \times \mathbf{v} \times \mathbf{B} - \frac{1}{R_m} \nabla \times \nabla \times \mathbf{B} \tag{4}$$

$$\nabla \cdot \mathbf{B} = 0 \tag{5}$$

supplemented by an equation of state,

$$p = \rho T, \tag{6}$$

where: $\rho(\mathbf{x}, t)$ is the mass density, $\mathbf{v}(\mathbf{x}, t) = (u, v)$ is the flow velocity, $P(\mathbf{x}, t)$ is the thermal pressure, $\mathbf{B}(\mathbf{x}, t) = (B_x, B_y)$ is the magnetic induction field, $T(\mathbf{x}, t)$ is the plasma temperature, $\zeta_{ij}(\mathbf{x}, t) = e_{ij} - \frac{2}{3} \nabla \cdot \mathbf{v} \, \delta_{ij}$ is the viscous stress tensor, $e_{ij} = (\partial_j v_i + \partial_i v_j)$ is the strain tensor and $\gamma$ is the adiabatic ratio. We use an extremely simplified diffusion model in which the thermal conductivity ($\kappa$), the magnetic resistivity ($\eta$) and the shear viscosity ($\mu$) are constant and uniform. For simplicity Stokes relationship is assumed, so the bulk viscosity $\lambda = (2/3)\mu$. The important dimensionless numbers are: $R = \rho_0 V_0 L_0/\mu \equiv$ Reynolds number, $R_m = V_0 L_0/\eta \equiv$ resistive Reynolds number, $Pr = C_p \mu/\kappa \equiv$ Prandtl number, $M = V_0/c_s \equiv$ free stream sonic Mach number, and $M_A = V_0/V_A \equiv$ Alfvénic Mach number. In these definitions, $\rho_0$ is a characteristic density, $V_A$ is a characteristic Alfvén speed, $L_0$ is a characteristic length, $C_p$ is the specific heat at constant pressure, $c_s$ is the sound speed, and $V_0$ is a characteristic flow speed. Time ($t$) is measured in units of fluid transit times ($= L_0/V_0$). In analogy with hydrodynamic problems, we refer to the spatial coordinate aligned with the mean flow as the streamwise direction ($x$), the spatial coordinate along which the mean flow varies as the cross-stream direction ($y$). The dimensions of the 2D numerical box are $L_x = 2\pi/\alpha$, where $\alpha$ is the streamwise wavenumber of the linearly unstable perturbation, and $L_y = \pm 50$, with the zero in $y$ centered in the box at the initial location of the velocity and magnetic shear layers.

## 2.2. Initial Conditions and Computational Method

We model a planar magnetic neutral sheet embedded in the center of a wake of transverse dimension $\Delta = L_0$, which we equate to the size of a GC molecular cloud (Shore & LaRosa 1999). The model is characterized by three parameters, namely the Alfvénic Mach number $M_A$, the sonic Mach number $M$, and the magnetic field amplification parameter $a_M$. This last quantity is the ratio of the internal flow amplified magnetic field to that advected by the ambient wind. The initial dynamical state is given by:

$$\hat{\mathbf{U}}_0(y) = (1 - \operatorname{sech} y)\hat{\mathbf{e}}_x, \quad \hat{\mathbf{B}}_0(y) = [(1 + a_M \operatorname{sech} y) \tanh y]\hat{\mathbf{e}}_x. \tag{7}$$

which are the velocity and magnetic field strength profile normalized to their values in the wind, $V_0$ and $B_\infty$. These profiles are shown in Figure 1. To ensure force balance we introduce a kinetic pressure profile which corresponds to the following normalized density and temperature profiles:

$$\rho = (1 + \beta_\infty - |\mathbf{B}|^2)/\beta_\infty, \quad T_0 = 1, \tag{8}$$

where $\beta_\infty = 2M_A^2/(\gamma M^2)$. In this paper we assume the following values for physical quantities at large distances in the cross-stream coordinates: a free stream velocity $V_0 = 2000$ km s$^{-1}$, a number density $n_\infty = 1$ cm$^{-3}$, and a magnetic field strength $B_\infty = 10^{-5}$ G, leading to an exterior Alfvén velocity $V_{A,\infty} = 6.3$ km/sec. It follows that $M_A = 317$ and $M = 5.4$, since the sound speed is $c_{s,\infty} = 370$ km/sec. The magnetic amplification parameter is chosen to be $a_M = 100$. This is about the value required by our previous analytic estimate and the numerical simulations of Gregori et al. (2000). For $L_0 \approx 1$ pc, the crossing time is about 500 yrs.

A linear code is used to determine the wavenumber of the fastest growing unstable mode, $\alpha_{\mathrm{max}}$. This length is used to normalize the length scale $L_x$ required for the full nonlinear calculation. The linear calculations are carried out with an MHD Chebyshev collocation code (see Dahlburg & Einaudi, 2000). The nonlinear calculations are performed with the Versatile Advection Code (VAC) (Toth 1996).

We disturb the equilibrium with a periodic cross-stream momentum perturbation to ensure a full development in wavenumber space. These perturbations take the following form:

$$f(x, y) = \epsilon(\sin \alpha x \tanh y + \cos \alpha x \cosh y) \exp(-y^2). \tag{9}$$

For the run described in this paper we set $\epsilon = 0.001$. Following DKE00, this $\epsilon$ is small enough to insure that the unstable modes are naturally excited and not forced. In general, there are three broad classes of unstable modes: sinuous, ideal varicose, and resistive varicose. The last of these is related to the perhaps more familiar tearing mode. Since we are mainly concerned with the disruptive effects of the Kelvin-Helmholtz (KH) instability for a low $\beta$ configuration, we initialize the calculations with the most unstable *sinuous* mode. As shown in DKE01, this is the fastest growing mode for a compressible wake. The dispersion relation is shown in Figure 2; the relevant linearized equations can be found in Dahlburg & Einaudi (2000). For the sinuous mode $\alpha_{max} = 0.46$ corresponding to a maximum growth rate of 0.0367. Note that this is in good agreement with previous results for the unmagnetized supersonic wake (*e.g.* Chen et al. 1989). In physical units, $\alpha = 0.46$ corresponds to a growth time of $8.6 \times 10^4$ yr. For the numerical simulation we present here we used $\alpha = 0.25$ (linear growth rate of 0.0326) in order to resolve both the length and time scales of the fastest growing (sinuous) mode. This nonlinear simulation is performed on a $200 \times 200$ mesh which is stretched in the cross stream direction to provide greater accuracy in the vicinity of the shear layer.

## 3. Results

Disruption in a low $\beta$ neutral sheet such as a coronal streamer within a wind is instigated by an MHD instability, the tearing mode. In our high $\beta$ case it is dynamical in origin – the Kelvin-Helmholtz instability – and results in strong coupling of the wake plasma to the background flow. This leads to plasmoid formation and their subsequent acceleration. The calculation proceeds in pseudo-Lagrangian fashion; *i.e.* it fixes the size of a computational box and follows the development of that parcel of gas in time. To translate this timescale into a position along the filament requires integrating the equation of motion for a fluid parcel as the central flow accelerates. The neutral sheet is initially at rest and only slowly accelerates. Figure 3a shows the time evolution of the central velocity and Fig. 3b shows the corresponding streamwise (axial) displacement of a fluid parcel. Note that the central section of the flow is accelerated up to a little over 80% of the free stream velocity. Figure 4 shows the streamwise velocity at different times for cross cuts in the $x$ direction. Notice that the wake has barely broadened before about 220 growth times and has expanded by a factor $> 5$ by 250 growth times. In other words, once the instability sets in and the wake material has reached its terminal velocity, the wake quickly blows itself apart. In physical units, the fluid has moved about 50 pc ($\approx$ 50 initial wake widths) by the end of the linear growth interval. The filament has completely disrupted within another 50 growth times which corresponds to an additional displacement of about 40 pc. We therefore predict from this simulation that the length of a typical isolated NTF should be about 50 to 100 times its width.

Figure 5 shows the time development of the perturbed kinetic energy. Here $E_{kin}$ has been computed by subtracting the mean streamwise velocity. The dashed line is a linear fit to the fully nonlinear results from the onset of linear growth to when the instability saturates. This fit gives a growth rate $\approx 0.03189$, which is in good agreement with the linear result.

There is enough energy in the magnetic field to produce all of the observed high energy electrons but it is the growth of the instability that drives the reconnection and generates the necessary MHD turbulence for particle acceleration. The necessary turbulent energy is proportional to $E_{kin}$. We also know from the magnetic field development that reconnection is occurring in this simulation. Previous simulations (*e.g.* Santillan et al. 1999; Gregori et al. 2000) have shown that a neutral sheet does indeed form. Further, Gregori et al. found the onset of field annihilation and structural changes that they interpreted as reconnection; these occurred, however, without the drastic disruption and shredding of the field that we find. Figure 6 shows various stages in the time (hence spatial) evolution of $B^2$. Note that the figure presents the development of the field structure following the formation of the current sheet so the actual amplification is only about a factor of 3 between the maximum in Figure 1 and that in the shredded field. From Fig. 4, the broadening of the wake at about 220 to 250 growth times and saturation of the instability correspond to the shredding stage seen in Fig. 6. The transition is quite abrupt and this breakdown should mark the observational end for the nonthermal emission from a filament. From Fig. 7, where we display $B^{3/2}\rho$, a quantity proportional to the synchrotron emissivity, notice that the peak emissivity should occur where the instability is growing nonlinearly and is well downstream from the cloud. The structure is almost identical to Fig. 6 except in those sites where ideal MHD (field freezing) is starting to break down and reconnection is occurring. In this regard, the present calculation differs from those in the literature in exhibiting reconnection instead of merely field strength reduction. Although a complete calculation of particle acceleration and propagation is needed for quantitative comparisons with observations, by choosing times appropriately even this Lagrangian simulation can be used to produce the pseudo-radio image that we show in Fig. 7. It is interesting that this simulation predicts a previously unexplained feature of the radio observations, that the brightness peaks more or less at the midpoints of the NTFs and not at either end (LaRosa et al. 2000).

## 4. Discussion

Most of the NTFs show no spatial variation in their spectral index. Our simulations appear to agree with this. Although the surface brightness should drop quickly at the end of the filament as the field shreds and the wake expands, there is initially (for almost 100 growth times) no decrease in the local magnetic field strength. This is shown in Figure 7. On the other hand, once the instability saturates, both the strength and filling factor of the field drop rapidly. We would expect, then, to see a rapid decrease in the synchrotron emissivity at all frequencies and, if rapid enough, to not see an associated change in spectrum. The NTF system G359.85+0.39 is an exception to the general spectral properties of the isolated filaments (LaRosa, Lazio, & Kassim 2001) and may be the one observed case that shows the shredding region. The field shredding may be related to another observed tendency of the isolated NTFs to show multiple strands. Quantitative comparisons are difficult because we have performed only a two dimensional simulation. Yet it is suggestive that the appearance of such structures once the instability begins to develop mimic those seen in the high resolution radio images and may be expected as well in fully three dimensional simulations.

Finally, the stellar distribution in the galactic center is not spherically symmetric. We consequently expect that the GC wind will also not necessarily flow radially at all locations above plane. The fact that all the filaments – with one notable exception, G358.85+0.47 – point roughly perpendicular to the galactic plane is consistent with our model. The one exceptional set of filaments is the farthest in projection from Sgr A, about 225 pc. Since the clouds have a more nearly radial velocity than the wind relative to the GC, it may not be surprising that at this distance the wind is aberrated relative to the cloud's motion.

## 5. Summary

In summary, our cometary scenario leads to an instability that couples the static wake material to the external fast wind. This coupling accelerates the wake to about 80% of the wind speed, broadens the wake, and shreds the original internal magnetic field into filaments and accelerates the resulting plasmoids. In combination, these processes signal the breakdown of structure and limit the length of the filament. We therefore identify the observed end of the NTFs with the location of this disruption which in our simulation occurs at a distance of about 50 to 100 filament widths (about 50 to 100 pc). Particle acceleration is a natural consequence as the instability is growing and reconnection begins. Nonetheless, such questions require detailed simulations of the acceleration and radiation processes. These issues can now be addressed within the context of the cometary model.

We thank the anonymous referee for several very helpful questions. RBD was supported by the National Aeronautics and Space Administration (NASA) SOHO Guest Investigator Program and the NASA Space Physics Theory Program. GE was supported by the Ministero Università Ricerca Scientifica Scientologica (MURST). TNL was also supported by a Master Scholarship grant from Kennesaw State University. SNS was supported by the IUSB Distinguished Research Award and an intercampus research grant from Indiana University and by NASA. Computations were performed on the DoD HPC Shared Resource Center ERDC SGI O3K under the DoD HPCM Program.

Fig. 1.— Initial velocity and magnetic field profile of the magnetized wake.

Fig. 2.— Dispersion relation for the sinuous mode.

Fig. 3.— $a$. Mean streamwise velocity for the center of the wake as a function of growth time; $b$. Displacement-time relation for the simulation in units of most unstable modal wavelength and growth time (see text).

Fig. 4.— Time evolution of the $x$-averaged streamwise velocity.

Fig. 5.— Perturbed kinetic energy $E_{\rm kin}$ vs time. The dashed line is the linear fit to the pre-saturation interval of spectral development.

Fig. 6.— Filled contour plots of $|\mathbf{B}|^2$. The grayscale is linear with white maximum and black minimum. The maximum amplification relative to the peak in Fig. 1 is about a factor of 3 in field strength. The times for each snapshot are 190, 220, 250, and 280 growth times corresponding to about 5, 20, 40, and 60 distance units (see text).

Fig. 7.— Filled contour plots of $B^{3/2}\rho$ for the same times as Fig. 6. The grayscale is again linear. Note the strongly enhanced region in the third panel. The regions where reconnection is taking place are seen by comparing this figure with Fig. 6; a frozen-in field would show identical structures while departures from ideal MHD are present in the third and fourth panels.

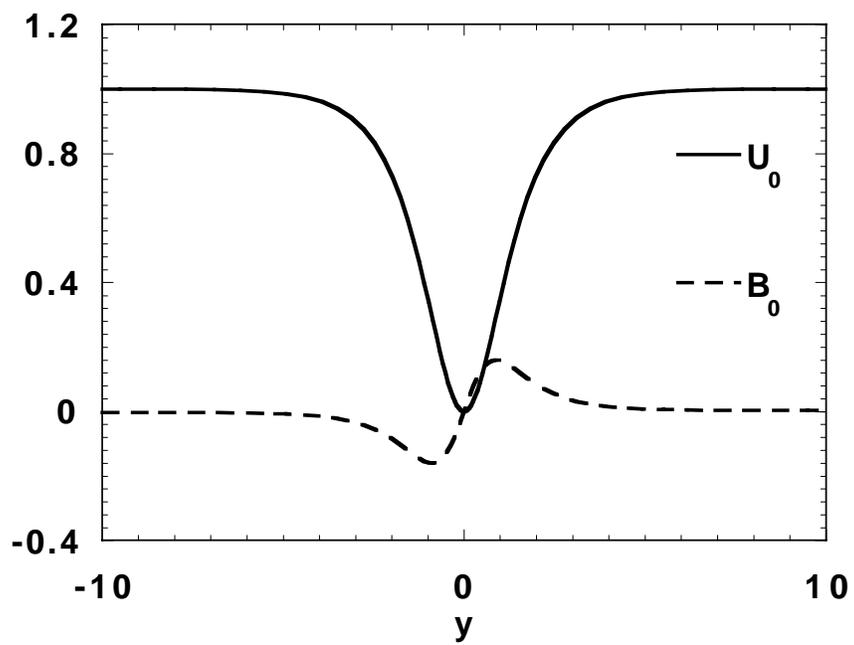

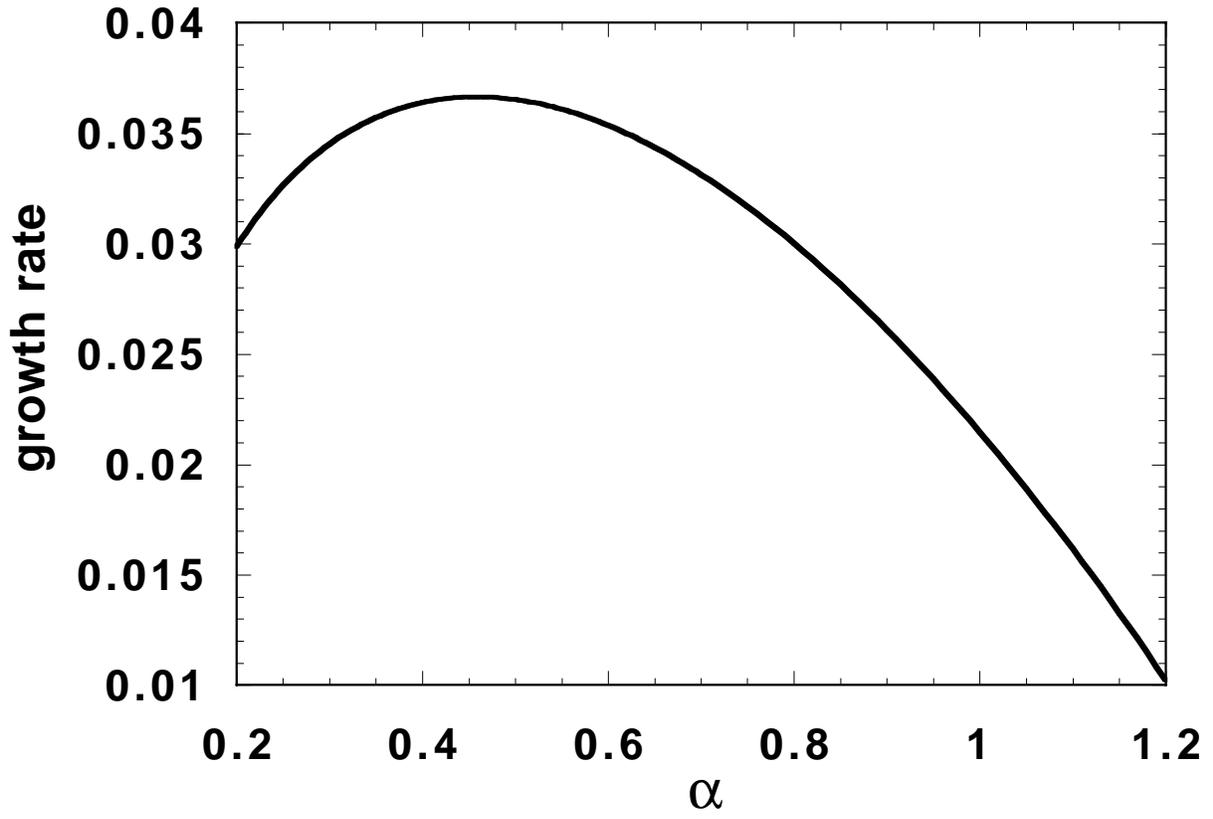

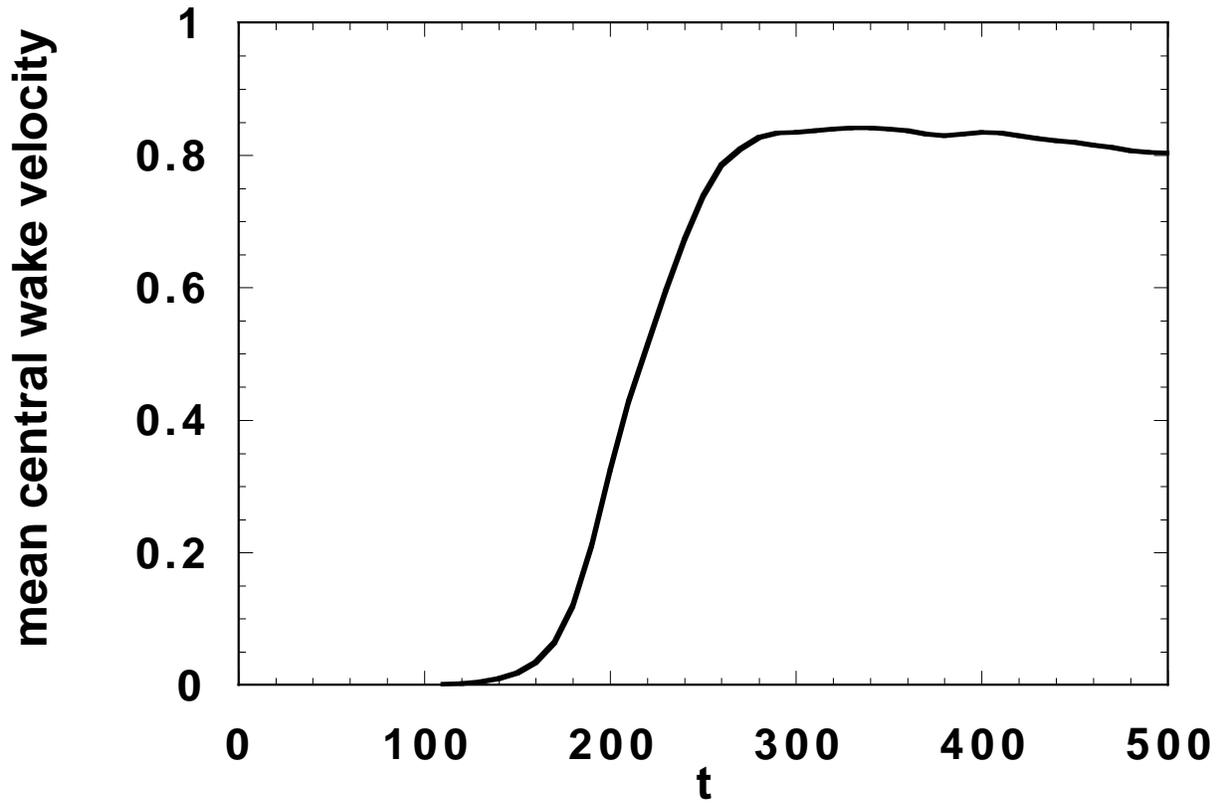

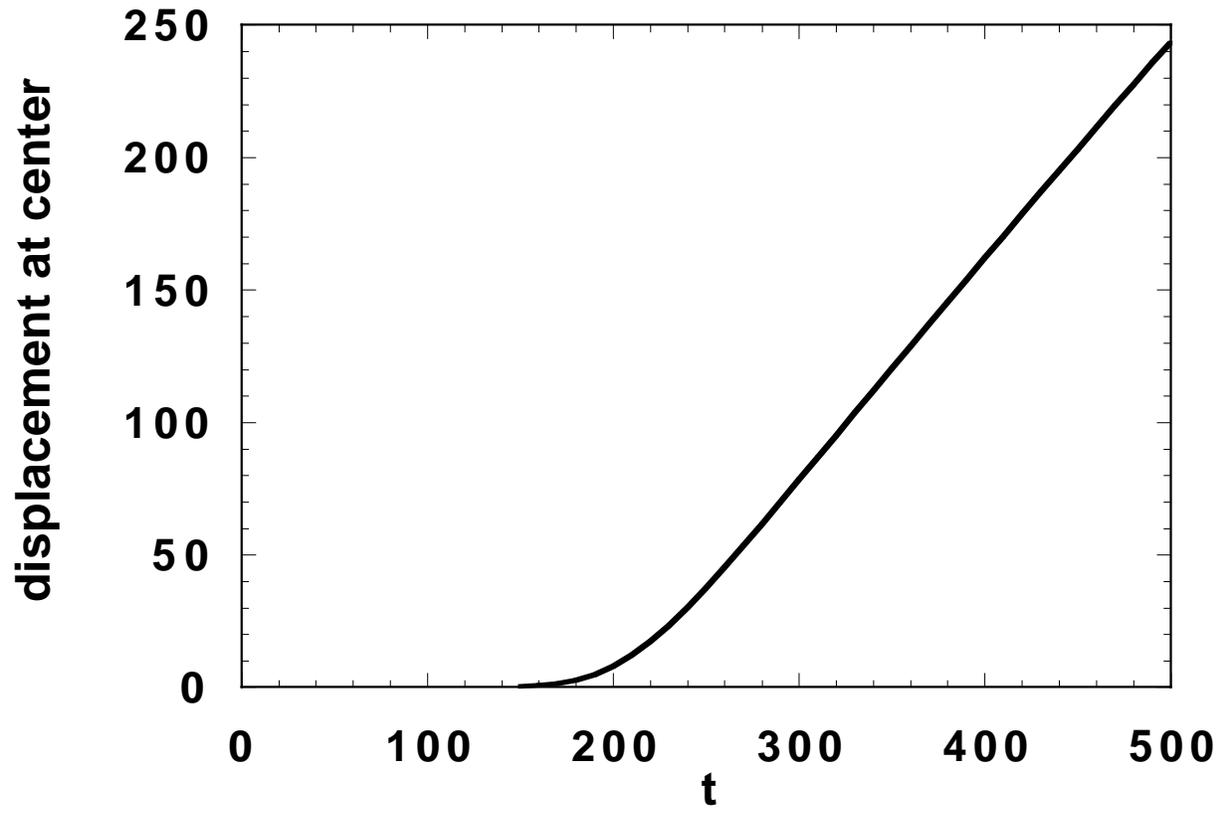

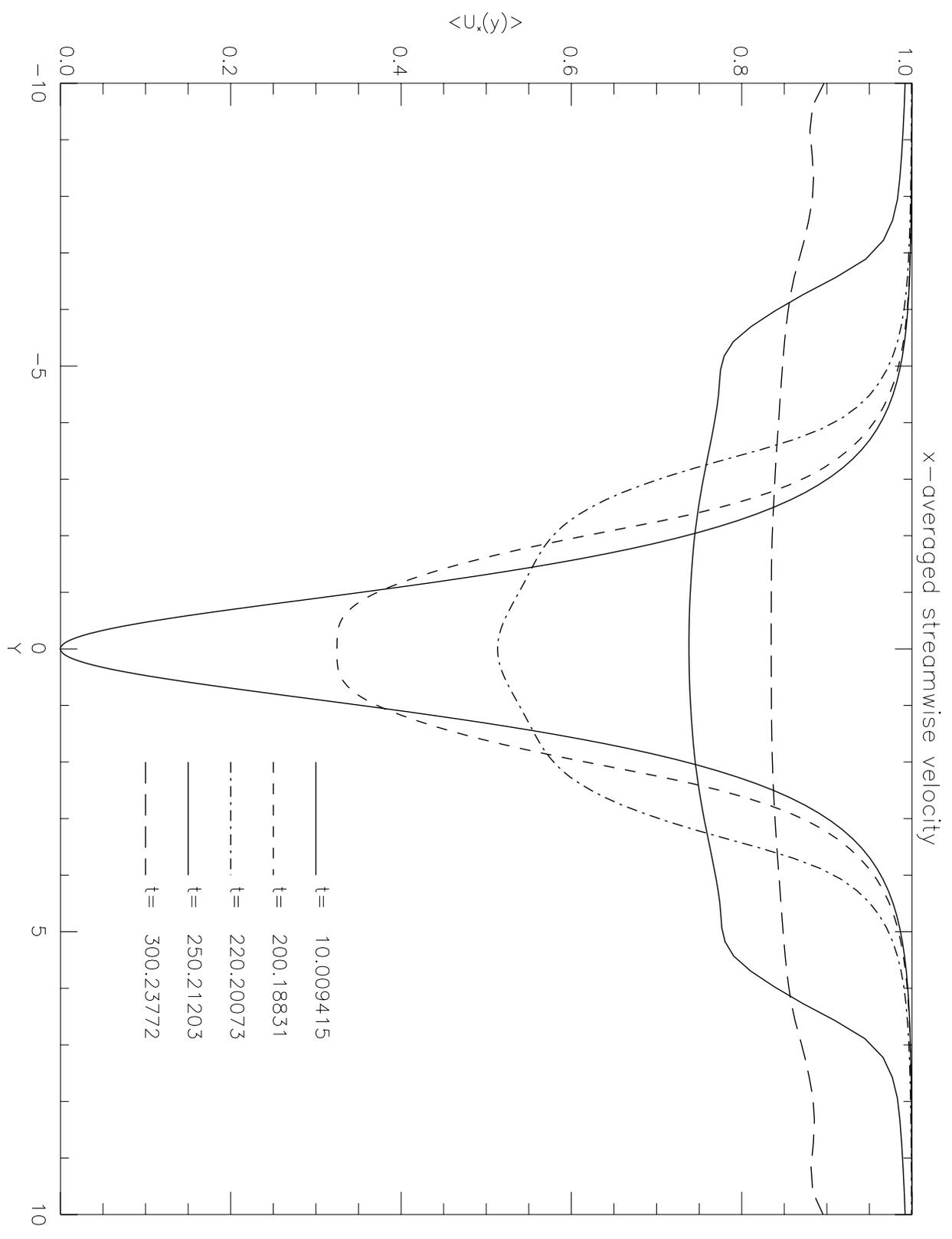

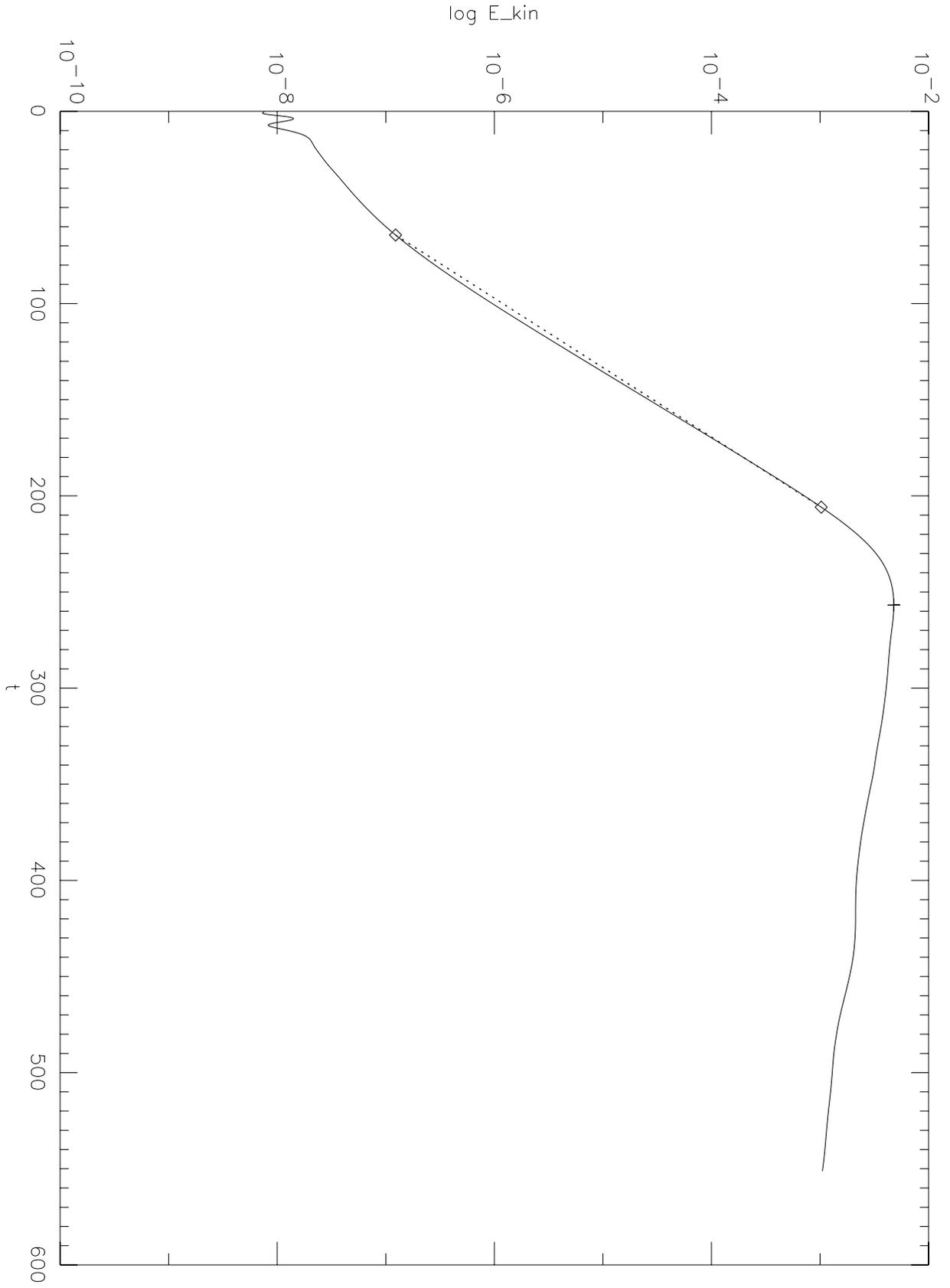

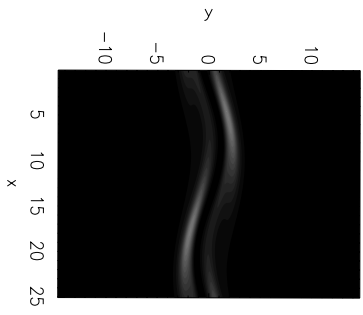

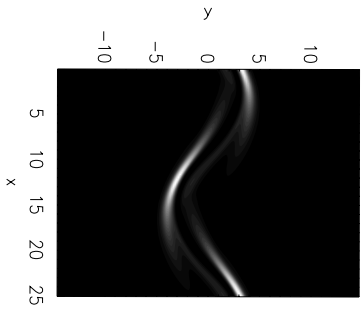

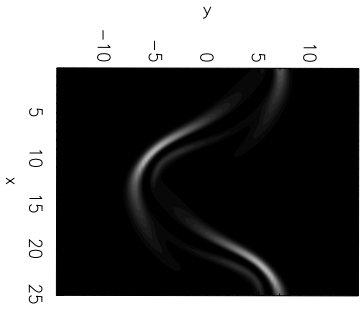

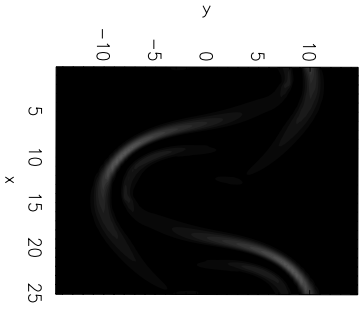

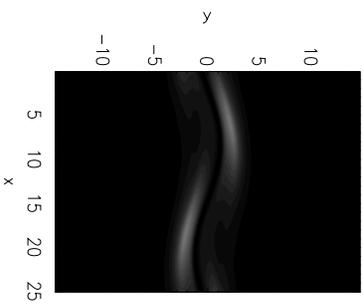

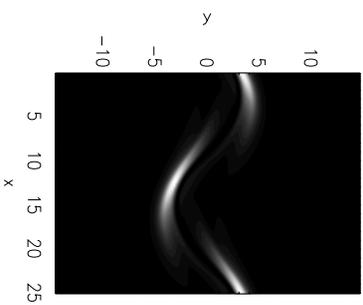

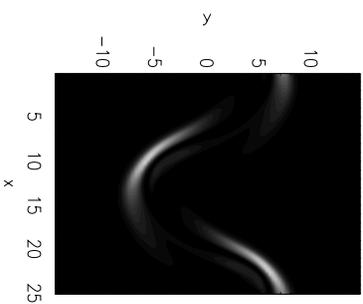

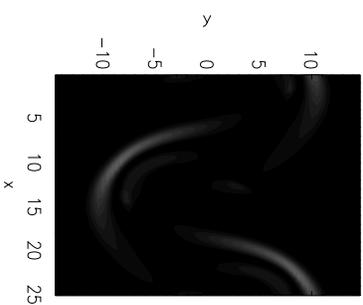